\documentclass[journal]{IEEEtran}

\ifCLASSINFOpdf
\else
\fi
\usepackage{amsmath,mathtools}

\DeclarePairedDelimiter\floor{\lfloor}{\rfloor}
\usepackage{algorithmic,algorithm}

\usepackage[caption=false,font=footnotesize,farskip=0pt,captionskip=0pt]{subfig}
\usepackage{hyperref}
\newlength\myindent
\newcommand\bindent{%
  \begingroup
  \setlength{\itemindent}{\myindent}
  \addtolength{\algorithmicindent}{\myindent}
}
\newcommand\eindent{\endgroup}

\usepackage{multirow}

\begin{document}
%
\title{Mitigating Sybil Attacks on Differential Privacy based Federated Learning}
%

%
%
\author{Yupeng Jiang,~\IEEEmembership{Member,~IEEE,} Yong Li,~\IEEEmembership{Member,~IEEE,} Yipeng Zhou,~\IEEEmembership{Member,~IEEE,}\\ and Xi Zheng,~\IEEEmembership{Member,~IEEE}}
\maketitle

\begin{abstract}
In federated learning, machine learning and deep learning models are trained globally on distributed devices. The state-of-the-art privacy-preserving technique in the context of federated learning is user-level differential privacy. However, such a mechanism is vulnerable to some specific model poisoning attacks such as Sybil attacks. A malicious adversary could create multiple fake clients or collude compromised devices in Sybil attacks to mount direct model updates manipulation. Recent works on novel defense against model poisoning attacks are difficult to detect Sybil attacks when differential privacy is utilized, as it masks clients' model updates with perturbation. In this work, we implement the first Sybil attacks on differential privacy based federated learning architectures and show their impacts on model convergence. We randomly compromise some clients by manipulating different noise levels reflected by the local privacy budget $\epsilon$ of differential privacy on the local model updates of these Sybil clients such that the global model convergence rates decrease or even leads to divergence. We apply our attacks to two recent aggregation defense mechanisms, called Krum and Trimmed Mean. Our evaluation results on the MNIST and CIFAR-10 datasets show that our attacks effectively slow down the convergence of the global models. We then propose a method to keep monitoring the average loss of all participants in each round for convergence anomaly detection and defend our Sybil attacks based on the prediction cost reported from each client. Our empirical study demonstrates that our defense approach effectively mitigates the impact of our Sybil attacks on model convergence.
\end{abstract}


%
\IEEEpeerreviewmaketitle

\section{Introduction}
%
%
%
%

\IEEEPARstart{M}{achine} learning and deep learning techniques have become prevailing in artificial intelligence. Along with this rise of application in clinical imaging for pathologic diagnosis and intelligent agriculture for crop monitoring, data privacy and security are the major concerns during machine learning training and test procedure. Learning from unbalanced and non-IID (not independently and identically distributed) data while preserving privacy, federated learning methods are proposed to train global models on distributed devices \cite{DBLP:conf/aistats/McMahanMRHA17, li2020privacy, zhang2020fenghuolun, zhang2020achieving}. In federated learning, each device or organization as a client has private data of training that is inaccessible to other clients and the server, which protects data privacy and data security.

In federated learning, learning models are shared globally.
However, such a distribution scale introduces practical challenges to reliability against various attacks towards the system. For example, this framework is vulnerable to model poisoning attacks \cite{alistarh2018byzantine, DBLP:conf/icml/BhagojiCMC19, fang2019local}, and it becomes even worse to some specific model poisoning attacks, called the Sybil attack \cite{douceur2002sybil}. In one study, Fung et al. \cite{fung2018mitigating} demonstrate that a deep learning network model in federated learning can be easily subverted by using the Sybil attack. In such attacks, the clients' model updates are tampered with a backdoor into the learned model, even if a small fraction of clients are compromised \cite{DBLP:conf/icml/BhagojiCMC19, bagdasaryan2018backdoor}.

As a defense technique in cryptography, differential privacy \cite{dwork2006calibrating} is utilized in federated learning to protect data confidentiality between the communications among server and multiple clients. By adding a certain distribution of random noise on each client's update locally or on the aggregated global model, user-level differential privacy \cite{DBLP:conf/iclr/McMahanRT018} can be achieved when training a global model. In this setting, any specific user’s data will not influence the behaviours of a trained model no matter it is used for training or not during the learning process. However, there have some weaknesses in these methods. One of which is that differential privacy requires that the scale of additive noises has to match the scale of parameters in the model updates to preserve utility. Therefore, it needs a comparatively larger number of clients or a larger value of privacy budget compared to that in the central setting \cite{kairouz2019advances}. Inspired by this limitation of user-level differential privacy which leaves more space for the attack, we introduce our Sybil attacks that aim to manipulate the updated parameters inside local models such that the aggregated model in the server has a high cost of the prediction indiscriminately for training examples, which makes the global model converges slowly, or even leads to divergence.

We perform the study on differential privacy based Sybil attacks in federated learning settings. A key challenge for the attacker is that how the compromised model updates from Sybil clients can obscure the aggregation defense rules in the uncompromised server to deviate the global model from its original prediction. To address the challenge, we apply different strategies to carefully craft local model updates of Sybil clients according to different aggregation rules by manipulating different noise levels reflected by the local privacy budget $\epsilon$ of differential privacy. Our goal of the attack is to introduce higher training costs in the global model when the server aggregates all the clients' updates, including our crafted parameters in each training round compared with the original one. Our attack intuition is the accumulated high cost during the learning process may slow down the convergence of the global model significantly.

Existing defenses against model poisoning attacks replace the mean aggregation rule in the central server with a Byzantine-resilient algorithm as the robust aggregator \cite{pillutla2019robust, su2016fault}. However, these defenses do not take into account the scenarios where differential privacy is applied in the federated learning model. Whereas differential privacy prevents data leakage, federated learning models are still susceptible to model poisoning attacks, especially to Sybil attacks. To address this challenge, we propose our defense method to defend against our Sybil attacks on differential privacy based federated learning settings. Our proposed defense excludes those client updates inducing high loss values of prediction on the global model where the cost is evaluated based on the loss report from each client. Empirical results show that our proposed defense mechanism effectively mitigates the impacts of our Sybil attacks on the convergence of the global model.

There are three main contributions in this paper as follows:
\begin{itemize}
\item We implement Sybil attacks on differential privacy based federated learning architectures and show their impacts on model convergence. 
\item We propose a method to detect and defend our Sybil attacks based on the prediction cost reported from each client. The key insight is that the poisoned model parameters from Sybil clients can be identified by their induced high loss values of prediction on the global model. 
\item We apply our attacks to two recent aggregation defense mechanisms. We also conduct an empirical study to illustrate that our proposed defense method effectively defends against our Sybil attacks.
\end{itemize}

This paper is organized as follows. Section \ref{Related Work} discusses the state-of-the-art attacks on federated learning and corresponding defenses. In Section \ref{assumption}, we introduce how to attack differential privacy based federated learning using Sybil attacks. We describe our attacks and defense solution in Section \ref{attack} and Section \ref{design}. Section \ref{evaluation} analyzes the evaluation results using two public datasets. We finally conclude our work and close with future research direction in Section \ref{conclusion}.

\section{Related Work} \label{Related Work}
\subsection{Attacks and Defenses on Federated Learning} \label{att_fl}

The most distinction between federated learning and centralized machine learning is that federated learning trains a model collaboratively across distributed client devices. Thus, it opens up new attack surfaces such that the adversary can manipulate the model updates sent back to the server \cite{baruch2019little}. This class of adversarial attacks is known as model poisoning attacks. Since the corrupted model updates can be arbitrary, model poisoning attacks are generally viewed as the most powerful and worst-case attacks, which is also referred to as Byzantine attacks \cite{DBLP:journals/toplas/LamportSP82}. Currently, Byzantine attacks mainly aim to degrade model performance or even make the global model unusable. Based on the goal of the adversary, these Byzantine attacks can be further classified as \textit{untargeted attacks} \cite{kairouz2019advances}. 
However, the limitation is that the attacker has to know about the aggregation rule in the server entirely. Otherwise, the attacks would be much less effective. Another category of model poisoning attack is \textit{targeted attacks}. The aim is that the trained model is modified in desired behaviour for the adversary, such as misclassification on some specific tasks. Bhagoji et al. \cite{DBLP:conf/icml/BhagojiCMC19} revealed that the learned global model could be poisoned to the misclassify targeted objective while preserving the classification accuracy of the trained model. Moreover, it needs only a small portion of client devices to be compromised for their targeted model poisoning attacks. However, it assumes that the server uses the accuracy of validation data to detect anomalous updates. In the training process, validation data are not accurate enough compared to training data due to the stochastic gradient descent (SGD) algorithm. Therefore, the effect of this kind of attack is limited without the appropriate assumptions. In the literature, data poisoning attacks are explored comprehensively. In data poisoning attacks, the adversary tampers the training dataset of clients by replacing labels or adding perturbations to the original data \cite{DBLP:conf/icml/KohL17, DBLP:conf/ndss/LiuMALZW018}.
Targeted attacks are also referred to as backdoor attacks \cite{chen2017targeted}, in which the performance of the global model on specific tasks is influenced by manipulating client data. In the work of targeted attacks \cite{DBLP:conf/icml/BhagojiCMC19} for model poisoning attacks, the final weights update sent back by the malicious client is learned from the auxiliary data. Therefore, we can see that although the model poisoning attacks are more powerful, it is enormously important to investigate data poisoning attacks for well understanding of the relation between them. Although some past work has explored untargeted data poisoning that reduces the accuracy of the global model notably using crafted training data \cite{DBLP:conf/icml/BiggioNL12}, research directions on targeted data poisoning attacks are dominant.

A common method to aggregate the local models is using the mean aggregation rule \cite{DBLP:conf/aistats/McMahanMRHA17}. However, this model averaging is susceptible to adversarial attacks and hardly provides privacy guarantees. Many works have explored Byzantine-resilient defense mechanisms for federated learning. Specifically, recent works propose various robust aggregation rules against both untargeted and targeted attacks. As the popular aggregation defense mechanisms, Krum \cite{blanchard2017machine} and Trimmed Mean \cite{DBLP:conf/icml/YinCRB18} are proposed to be robust under untargeted adversarial settings. These methods replace the mean aggregation rule in the central server with a Byzantine-resilient algorithm as the robust aggregator. However, these mechanisms work under appropriate assumptions that provably asymptotic on the number of the client. With data poisoning attacks, they can be viewed as special cases of model poisoning attacks. The reason is that compromised training data will induce anomaly in clients' model updates. Therefore, Byzantine-resilient defenses against model poisoning attacks may also work for data poisoning attacks \cite{xie2019zeno}. It is noteworthy that any proposed robust defense has to guarantee the convergence of the global model when using a gradient descent algorithm on the client.

\subsection{Differential Privacy based Federated Learning}
Concerning about data privacy, user-level differential privacy \cite{DBLP:conf/iclr/McMahanRT018} is leveraged in the context of federated learning. Several works have shown that the use of differential privacy effectively defends against privacy disclosures on the scope of targeted model poisoning attacks \cite{sun2019can}, data poisoning attacks \cite{geyer2017differentially, DBLP:conf/ijcai/Ma0H19}, and attacks on adversarial examples \cite{lecuyer2019certified}. In our study, we implement the first Sybil attacks as untargeted model poisoning attacks and defenses on federated learning models with differential privacy applied.
In \cite{sun2019can}, Sun et al. have explored an approach to eliminate the impacts of targeted attacks using differential privacy, while our Sybil attacks are untargeted attacks that focus on differential privacy based federated learning.

Our proposed Sybil attacks are one of the worst-case model poisoning attacks. In Sybil attacks, an adversary is capable of manipulating model updates from a large number of clients, which significantly influences the performance of the global model, even when state-of-the-art defense aggregation rules are present. Furthermore, while recent Byzantine model poisoning attacks studies \cite{baruch2019little, DBLP:journals/toplas/LamportSP82} focus on Byzantine-robust federated learning, our attacks consider differential privacy preserved federated learning.

\section{Threat Model} \label{assumption}
This section describes the architecture of our differential privacy based federated learning settings, followed by characterizing the capabilities and goals of adversaries in Sybil attacks.

In this paper, we consider a standard federated learning context, in which there are $K$ clients in total, each owning private training data and the number of $c$ compromised clients at most from $K$ clients. All the clients collaboratively train a classifier by solving the optimization problem
\begin{equation}\label{op}
\min f(w) \qquad \text{where} \qquad f(w) = \sum_{k=1}^{K}f_{k}(w)
\end{equation}
where $f_{k}(w)$ is the objective function for the local dataset on the $k$th client, and $w$ denotes the parameters of the global model. The server aggregates clients' models by a predetermined aggregation rule $w=\mathcal{A}(w_1,w_2,\cdots,w_K)$ where $w_k$ denotes the parameters in the local model updates of each client.

We call our federated learning settings \textit{differential privacy based federated learning}. In this architecture, the clients' model updates are masked with a user-level differential privacy perturbation in the form:
\begin{equation}
\widetilde w_k=w_k+\widetilde n_k
\end{equation}
where $w_{k}$ is the parameters in the local model updates from the $k$th client, and $\widetilde n_k$ is an additive noise to guarantee differential privacy. The noise level of each client is reflected by the local privacy budget $\epsilon$ of differential privacy on the local model updates. In this paper, we use Laplace distributed additive noises with the following probability density function:
\begin{equation}\label{fllp}
f(w)=\frac{1}{2b}\exp\left(-\frac{\left\| w \right\|_1}{b}\right)
\end{equation}
where $w$ represents model parameters shared globally and $b$ is a scale parameter. In accordance with Theorem 1 in work \cite{wu2020value}, we calculate the scale parameter in \eqref{fllp} $b=2D_{max}T/n_k\epsilon$ where $D_{max}$ represents the maximum absolute value of the model update parameters in the current communication round, $T$ is the total number of communication rounds, and $n_k$ is the number of examples in training dataset of the $k$th client. It is proved that the clients' model updates being sent back to the server that perturbed by noise $\widetilde n$ with this Laplace mechanism meet $(\epsilon, 0)$ - user-level differential privacy.

Our Sybil attack is one of the untargeted model poisoning attacks. In this attack model, an adversary can spoof up to the number of $c$ clients and tamper model parameters before sending them back to the server during the training process, as the capability of the Byzantine threat model as discussed in Section \ref{att_fl}. In this paper, we assume that the $c$ compromised clients are from a total of $K$ clients and no more fake clients in the system for simplicity. Moreover, we assume that the adversary knows the aggregation rule $\mathcal{A}$ used by the server, as it is usually published for the trust and transparency of the system \cite{DBLP:conf/aistats/McMahanMRHA17}.


We consider the goal of an adversary is to slow down the convergence rates of the global model or even diverge the model in the training phase. In this work, we assume the loss function of models is smooth and strongly convex. Although the loss function for high-dimensional networks is usually non-convex, it can still achieve a local minimum using stochastic gradient descent (SGD) algorithm iteratively on each client when training a model in federated learning. Under this assumption, the lower bound of global convergence rate in federated learning settings can reach to $O(\frac{1}{T})$, where $T$ denotes the total number of communication rounds to train a model \cite{grimmer2019convergence}.

In the Laplace differential privacy mechanism, for any $\epsilon > 0$, the scale of additive noise is increased when $\epsilon$ is reduced, corresponding to an increased level of privacy protection. Therefore, differential privacy with a small value of $\epsilon$ reduces the convergence rate of the global model. Given the lower bound of the model convergence rate, we propose a search algorithm to choose an optimal $\epsilon$ for differential privacy in the system. Specifically, we select one from the following values as the local privacy budget $\epsilon$ of clients: 0.1, 0.3, 0.5, 1.0, 2.0, 5.0, 8.0, 10.0. These typical values of $\epsilon$ have been evaluated in recent works for the trade-off investigation between privacy and utility of differential privacy \cite{abadi2016deep, wu2020value, zhu2019deep}. We first initialize $\epsilon = 10.0$ for all clients and calculate the global convergence rate based on the average loss of prediction on the global model in 50 iterations of training. If it is greater than $O(\frac{1}{T})$, then we choose the next smaller value of $\epsilon$ from candidates and repeat this process until the global convergence rate is less than $O(\frac{1}{T})$. This process determines the optimal value of $\epsilon$ for honest clients, which guarantees differential privacy, meanwhile preserves the convergence rates of the federated learning model. To solve $\epsilon$ value on Sybil clients, we will introduce our attack strategies according to different aggregation rules respectively in the next section.

\section{Our Attack} \label{attack}
In the user-level differentially privacy-preserving federated learning setting, for any $\epsilon > 0$, the scale of additive noise over the client updates is increased when $\epsilon$ is reduced. We leverage this characteristic to introduce a larger variance on model updates from Sybil clients using a smaller value of $\epsilon$ relative to it on honest clients, which will induce a higher loss of prediction on the global model in each iteration round of training.

In this section, we introduce our Sybil attack strategies for three aggregation rules in the central server of federated learning, including one widely used aggregator and two state-of-the-art defensive mechanisms.

\subsection{Our Attack to FedAvg}
One commonly used aggregation rule in federated learning is FederatedAveraging (FedAvg) \cite{DBLP:conf/aistats/McMahanMRHA17}. In the FedAvg algorithm, the global model in each communication round of training is the average of all clients' model parameters.
\begin{equation}\label{FedAvg}
w_{t+1}=\frac{1}{K}\sum_{k=1}^{K}w_{t}^{(k)}
\end{equation}
where $w_{t}^{(k)}$ represents the local parameters from the $k$th client in the current round $t$, and $w_{t+1}$ is the aggregated global model for the next training round. The user-level differential privacy on each client's model update is applied by:
\begin{equation}\label{DP}
\widetilde w_{t}^{(k)}=w_{t}^{(k)}+\widetilde n_{t}^{(k)}
\end{equation}
where $\widetilde n_{t}^{(k)}$ is a Laplace additive noise $\sim\mathcal{L}(0, \frac{\Delta f}{\epsilon})$ with the optimal $\epsilon$ we choose using our search algorithm. Based on this optimal $\epsilon$, the differential privacy can be guaranteed while preserving the convergence rate of the global model above the lower bound. From \eqref{FedAvg}, it is easy to get the aggregation with differential privacy:
\begin{equation}\label{FedAvgDP}
\widetilde w_{t+1}=\frac{1}{K}\left(\sum_{k=1}^{K}w_{t}^{(k)}+\sum_{k=1}^{K}\widetilde n_{t}^{(k)}\right)
\end{equation}
As discussed in Section \ref{att_fl}, FedAvg is vulnerable to adversarial attacks. We can simply increase additive noise $\widetilde n_{t}^{(k)}$ in \eqref{FedAvgDP} from Sybil clients to achieve a large variance on $\widetilde w_{t+1}$ by reducing the local privacy budget $\epsilon$ on Sybil clients.

We propose two strategies to attack FedAvg. One obvious method is to use any $\epsilon_s$ on Sybil clients for $0 < \epsilon_s < \epsilon_h$ where $\epsilon_h$ is the optimal value we choose for honest clients using our search algorithm in Section \ref{assumption}. In this work, for simplicity, we assume that all Sybil attackers use the same $\epsilon_s$ for attacks, and all honest clients use the same $\epsilon_h$ for differential privacy. The smaller value of $\epsilon_s$ corresponds to the stronger attack intensity, easier to be detected, however. To choose the value of $\epsilon_s$ from candidates, we evaluate this attack method using different $\epsilon_s$ and the different number of Sybil attackers. The other more stealthy method is using synchronous additive noises on these collusive Sybil clients. In this method, the noises added to the model updates of Sybil clients are from either the positive or negative part of Laplace distribution in phase. According to \eqref{FedAvgDP}, even with a small magnitude of additive noises, the attack intensity will be amplified by the sum operation of FedAvg when the server aggregates all the clients' model updates. We evaluate two methods in our experiment, respectively.

\subsection{Our Attack to Krum}
Recent work \cite{blanchard2017machine} proposed Krum to increase the robustness of the aggregation rule against Byzantine attacks. The basic idea is that it selects one of the model updates from all the clients as the global model instead of using the mean of them. The selection criterion is based on the similarity concerning Euclidean distance between two clients' model updates. Specifically, suppose we have $K$ clients in total and $c$ Sybil clients among them, it first calculates the Euclidean distance between each client's model update. Then for each model update, it computes the squared sum of the smallest $K-c-2$ Euclidean distances. Finally, the Krum algorithm selects the model update with the minimum squared sum as the global model. It has been proved that the global model can converge to a local minimum under Byzantine attacks when $c<\frac{K-2}{2}$ by using Krum. This literature also proposed the Multi-Krum algorithm as a variant version of Krum to speed up the convergence when training a global model. In Multi-Krum, it selects $m$ clients' model updates with the smallest squared sum instead of one in Krum, then uses the mean of selected model updates as the global model. We can see that when $m=1$, Multi-Krum is the same as Krum, and when $m=K$, Multi-Krum is the FedAvg aggregation rule.

As Krum selects one model update from $K$ clients as the global model for the next communication round, our idea is that this model is from one of $c$ Sybil clients. The goal is this selected model deviates the global model from its intended converge direction before attacks. The key challenge of the attack is that each crafted local model with added random noises will induce a large Euclidean distance to the models from honest clients. As a result, Krum can easily exclude our crafted local models in such an aggregation rule. To address this challenge, we let model updates from Sybil clients maintaining the same to achieve a zero Euclidean distance between each of them. Then we carefully adjust $\epsilon_s$ in these Sybil clients such that their Euclidean distances to honest models are comparable with those among honest clients. This collusion of Sybil clients ensures our crafted model update to be selected by Krum. In implementing our attack, we evaluate both Krum and Multi-Krum on different training models respectively to maximize the attack impacts in experiments.

\subsection{Our Attack to Trimmed Mean}
Another aggregation rule, Trimmed Mean \cite{DBLP:conf/icml/YinCRB18}, considers the element-wise algorithm in the model updates. Similar to Krum, Trimmed Mean requires an explicit number of compromised clients. As we assumed those $c$ Sybil clients as mentioned above, it removes the largest and smallest $c$ elements in model parameters among all clients' updates. After that, it uses the average of the remaining elements as the corresponding parameter in the global model. In Trimmed Mean, the variance of model parameters in clients' update is constrained to a benign magnitude, which mitigates the impacts of Byzantine attacks. The authors also proved that the global model converges when $c<\frac{K}{2}$ and the statistical error rates achieves $O(\frac{c}{K\sqrt{n}}+\frac{1}{\sqrt{Kn}})$ for strongly convex loss functions, where $n$ is the number of examples in the training dataset of each client. We notice that when $c=0$, i.e. there is no attack, the Trimmed Mean algorithm is equivalent to FedAvg.

To slow down the convergence of the global model in our attack, we craft $c$ compromised local models based on the intended gradient of each element in the current training round. Specifically, when one parameter in the global model intends to increase upon the previous iteration if there is no attack, we add negative random noise with Laplace distribution onto this element of the corresponding location in each compromised client, such that this parameter with additive noise from each compromised client is smaller than the majority of the corresponding model parameter from the honest clients. As a result, the mean of the remaining $K-2c$ elements according to the Trimmed Mean algorithm is tending to decrease upon the previous iteration. Otherwise, if one parameter in the global model intends to decrease upon the previous iteration, we add positive random noises with Laplace distribution on each compromised client in the same way. In our experiments, we evaluate the values of $\epsilon_s$ to nominate the most effective attacks.

\section{Our Defense} \label{design}
We design a method to detect and defend our Sybil attacks on differential privacy based federated learning setting. Compared with state-of-the-art defense mechanisms such as Krum and Trimmed Mean, our proposed algorithm does not require the exact quantity of compromised clients. The experiment results show that our defense method effectively mitigates the impacts of untargeted model poisoning attacks on model convergence.

The key insight is that the poisoned model updates from Sybil clients can be identified by their induced high loss of prediction on the global model. Technically, the cost of a network is defined as a function $f_i(w)=\ell(x_i,y_i;w)$ that takes model parameters $w$ as its input and maps the loss of output on examples $(x_i,y_i)$ where $x_i$ is input and $y_i$ is the label. In our federated learning settings, Sybil clients will contribute model updates that appear larger loss values than those from honest clients to affect the convergence of the global model.

Our approach keeps monitoring the convergence rate of the global model throughout all training rounds on the server-side. As we discussed in Section \ref{assumption}, the optimal convergence rate can reach to $O(\frac{1}{T})$ for a smooth and strongly convex loss function where $T$ denotes the number of communication rounds. Our algorithm evaluates the convergence rate from round 2 by comparing the model loss decrease rate to a pre-determined threshold to detect Sybil attacks. This threshold reflects defense intensity. For most of the machine learning or deep learning models, the convergence rate is usually ranging from $O(\frac{1}{\sqrt{T}})$ to $O(\frac{1}{T})$ \cite{grimmer2019convergence}. In this paper, we set the threshold to $0.8(\frac{1}{t-1}-\frac{1}{t})$ or $0.8(\frac{1}{\sqrt{t-1}}-\frac{1}{\sqrt{t}})$ depending on the loss function in the global model. Note that a ratio of 0.8 is used to tolerate non-malicious failures from unreliable clients.

To locate Sybil attackers from participating clients, we use binary search in the vector of clients. For each half of clients, the central server sends loss report requests to a random fraction $C$ of $K$ clients with model parameters $w$ averaged from model updates in the corresponding half of clients. After that, the loss values reported from selected clients $l^k$ are averaged in the server. We keep searching Sybil attackers in the half of client vector with a larger mean of loss until three clients or 10\% of total $K$ clients remaining. Finally, we aggregate all clients’ model parameters, excluding these remaining client updates after the binary search has finished for the next global training round. Although there could have some honest client updates sacrificed for a few rounds of communication, our defense does not influence the convergence of the global model. The details of our defense method are introduced in Algorithm \ref{alg1}.

\begin{algorithm}
\caption{Detection and defense}\label{alg1}
\begin{algorithmic}
\REQUIRE Average training loss $l_{1}$ in round $1$
\bindent
\STATE \textbf{Server executes:}
\eindent
\FOR{round $t=2,3,...$}
  \IF{$(\Delta l_{t})/l_{1}<$ threshold of defense intensity}
  \STATE // \textit{The binary search}
  \STATE // \textit{The invariant: the average loss of clients containing Sybil attackers is larger than that of the other half.}
  \STATE $i,h=1,K$
  \WHILE{$h-i>\max(K/10,2)$}
    \STATE $m=\floor{(i+h)/2}$
    \STATE $w_{t}^{\prime}=\frac{1}{m}\sum_{k=1}^{m}w_{t}^{(k)}$
    \STATE $w_{t}^{\prime\prime}=\frac{1}{K-m}\sum_{k=m+1}^{K}w_{t}^{(k)}$
    \STATE // \textit{Each client reports loss with model parameters $w_{t}^{\prime}$ and $w_{t}^{\prime\prime}$}
    \STATE $S_{t}=$ random set of $\max(C\cdot K,1)$ clients
    \FOR{each client $k\in S_{t}$}
      \STATE $l_{t}^{(k)\prime}=$ ClientCost$(k,w_{t}^{\prime})$
      \STATE $l_{t}^{(k)\prime\prime}=$ ClientCost$(k,w_{t}^{\prime\prime})$
    \ENDFOR
    \STATE $l_{t}^{\prime}=\frac{1}{num(S_{t})}\sum_{k=1}^{num(S_{t})}l_{t}^{(k)\prime}$
    \STATE $l_{t}^{\prime\prime}=\frac{1}{num(S_{t})}\sum_{k=1}^{num(S_{t})}l_{t}^{(k)\prime\prime}$
    \IF{$l_{t}^{\prime}<l_{t}^{\prime\prime}$}
      \STATE $i=m+1$
    \ELSE
      \STATE $h=m$
    \ENDIF
  \ENDWHILE
  \STATE Exclude client updates ranging from $w_{t}^{(i)}$ to $w_{t}^{(h)}$, remaining $K'$ clients
  \ENDIF
  \STATE $w_{t+1}=\frac{1}{K'}\sum_{k=1}^{K'}w_{t}^{(k)}$
\ENDFOR
\STATE
\bindent
\STATE \textbf{ClientCost($k,w$):}
\eindent
\STATE batches $\leftarrow$ training data split into batches of size $B$
\FOR{batch $b$ in batches}
  \STATE $l_{b}=\ell(w;b)$
\ENDFOR
\STATE $l=\frac{1}{num(batches)}\sum_{b=1}^{num(batches)}l_{b}$
\RETURN $l$ to server
\end{algorithmic}
\end{algorithm}

In Algorithm \ref{alg1}, our proposed defense method keeps monitoring the convergence rate of the global model from iteration round 2 until the end of the training period, by comparing the rate to a pre-determined threshold of defense intensity. When this model loss decrease rate drops below the threshold in a certain round $t$, our method detects Sybil attacks and launches the defense procedure. The binary search algorithm is used to locate Sybil model updates among all the participating clients. The invariant in binary search is the key insight of our proposed defense method, which is that the average loss of clients containing Sybil attackers is larger than that of the other half. Specifically, we split the model updates from all the clients and average the local model parameters of each half, $w_{t}^{\prime}$ and $w_{t}^{\prime\prime}$ respectively. Then the server sends both global model parameters $w_{t}^{\prime}$ and $w_{t}^{\prime\prime}$ to a randomly selected set of clients $S_{t}$ for the request of the loss report. Each client $k\in S_{t}$ completes the training task based on both global model parameters $w_{t}^{\prime}$ and $w_{t}^{\prime\prime}$ and returns loss value $l_{t}^{(k)\prime}$ and $l_{t}^{(k)\prime\prime}$ to the server respectively. The client uses the batches of size $B$ on the SGD algorithm locally. After all the clients in $S_{t}$ report loss values $l_{t}^{(k)\prime}$ and $l_{t}^{(k)\prime\prime}$, the server calculates the mean of these loss values $l_{t}^{\prime}$ and $l_{t}^{\prime\prime}$ for each half client vector correspondingly. We keep searching for Sybil clients in half of the clients' vector with a larger loss value by comparing $l_{t}^{\prime}$ and $l_{t}^{\prime\prime}$. We repeat this process until there are 10\% of total $K$ clients or 3 clients left in the vector. Then we exclude these clients' model updates and average the remaining model updates as the global model $w_{t+1}$ for the next training round. This concludes our proposed defense mechanism against Sybil attacks for monitoring and detection implementation.

\section{Evaluation} \label{evaluation}
\subsection{Empirical Setting}
We consider a real-world scenario for our experiments. The hospital or pathological laboratory as the clients collaboratively train a model for pathologic diagnosis using their private clinical images of patients. A trusted third-party service provider as the central server is to complete aggregation operation. An adversary can manipulate the model updates from compromised clients.

In this paper, our proposed attack and defense approaches are evaluated by CNN and MLP models on two datasets MNIST and CIFAR-10 respectively. The MNIST data are partitioned by non-IID, and CIFAR-10 data are IID. We implement a federated learning prototype of PyTorch based on \cite{DBLP:conf/aistats/McMahanMRHA17}. The computer environment is Intel\textregistered ~Core\texttrademark ~i7-4770 CPU @ 3.40GHz processor, 16.0GB RAM and Windows 10 64-bit operating system.

The parameters for our differential privacy based federated learning settings by default are summarised in Table \ref{tab:fl_set}. The Sybil attackers are randomly selected from 100 clients in each communication round. All additive noises are Laplace distributed with a corresponding privacy budget.

\begin{table}
    \centering
    \caption{Federated Learning Settings}
    \label{tab:fl_set}
    \begin{tabular}{|c|c|c|}
        \hline
        \textbf{Parameter} & \textbf{Description} & \textbf{Value} \\
        \hline
        $K$ & Number of clients & 100 \\
        \hline
        $C$ & Fraction of clients & 0.1 \\
        \hline
        $c$ & Number of compromised clients & 20 \\
        \hline
        $T$ & Number of communication rounds & 50 \\
        \hline
        $B$ & Local batch size & 10 \\
        \hline
        $E$ & Number of local epochs & 5 \\
        \hline
        $\eta$ & Learning rate & 0.01 \\
        \hline
        $\epsilon_h$ & Privacy budget of honest clients & 8.0 \\
        \hline
        $\epsilon_s$ & Privacy budget of Sybil clients & 0.3 \\
        \hline
    \end{tabular}
\end{table}

To compare with our proposed attack methods, we implement the Gaussian attack \cite{fang2019local} as a benchmark. This attack injects random noises with Gaussian distribution into the local model updates from Sybil clients. In our experiments, we set the mean of the distribution to 0 and its standard deviation to 0.3 for evaluation.

As shown in Table \ref{tab:fl_set}, we perform 50 rounds of training and 5 local epochs on each client in our experiments. The learning rates were tuned between 0.01 to 0.05 for the best performance.

We use the error rate as test metrics for the evaluation of our defense, which is defined in \eqref{er}.
\begin{equation}\label{er}
Error~Rate = 1 - Testing~Accuracy
\end{equation}

\subsection{Evaluation on Our Attacks}
The empirical results for our attacks are shown in Table \ref{tab:attack_cnn} and Table \ref{tab:attack_mlp}. We replicate \textit{FedAvg}, \textit{Krum}, and \textit{Trimmed Mean} algorithms. We then apply the Gaussian attack and our proposed attack to these methods respectively. These results illustrate that the error rates after our proposed attacks are significantly higher than those after Gaussian attacks on both CNN and MLP models.

Firstly, Table \ref{tab:attack_cnn} shows the error rates after attacks on the CNN model as the classifier using the MNIST dataset. When there is no attack, the trained models achieve a low error rate of 3\% for all the aggregation rules. Krum and Trimmed Mean are robust to the Gaussian attack, where the error rates remain almost at 3\%. However, the error rates increase dramatically to 14\% and 85\% when we use our proposed methods to attack Krum and Trimmed Mean respectively.

Secondly, we notice that the FedAvg aggregator hardly defends against adversarial attacks, which need to be replaced with a robust aggregation rule. In Table \ref{tab:attack_cnn} and Table \ref{tab:attack_mlp}, they both show that the error rates of trained models increase significantly when we apply the Gaussian attack and our proposed attack to FedAvg.

\begin{table}
    \centering
    \caption{Error Rates on CNN Model After Attacks}
    \label{tab:attack_cnn}
    \begin{tabular}{|c|c|c|c|}
        \hline
        & No Attack & Gaussian Attack & \textbf{Proposed Attack} \\
        \hline
        FedAvg & 0.03 & 0.24 & \textbf{0.90} \\
        \hline
        Krum & 0.03 & 0.03 & \textbf{0.14} \\
        \hline
        Trimmed Mean & 0.03 & 0.05 & \textbf{0.85} \\
        \hline
    \end{tabular}
\end{table}
\begin{table}
    \centering
    \caption{Error Rates on MLP Model After Attacks}
    \label{tab:attack_mlp}
    \begin{tabular}{|c|c|c|c|}
        \hline
        & No Attack & Gaussian Attack & \textbf{Proposed Attack} \\
        \hline
        FedAvg & 0.59 & 0.73 & \textbf{0.91} \\
        \hline
        Krum & 0.59 & 0.61 & \textbf{0.63} \\
        \hline
        Trimmed Mean & 0.59 & 0.59 & \textbf{0.65} \\
        \hline
    \end{tabular}
\end{table}

\begin{figure}
    \centering
    \subfloat[FedAvg]{\includegraphics[width=7.5cm]{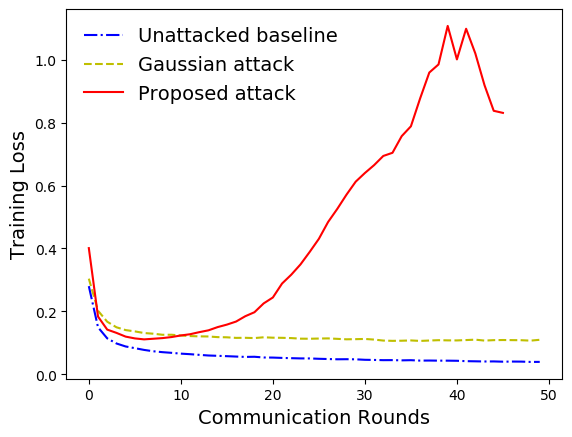}
    \label{fig:attack_avg_cnn}}
    \vspace{0.27cm}
    \subfloat[Krum]{\includegraphics[width=7.5cm]{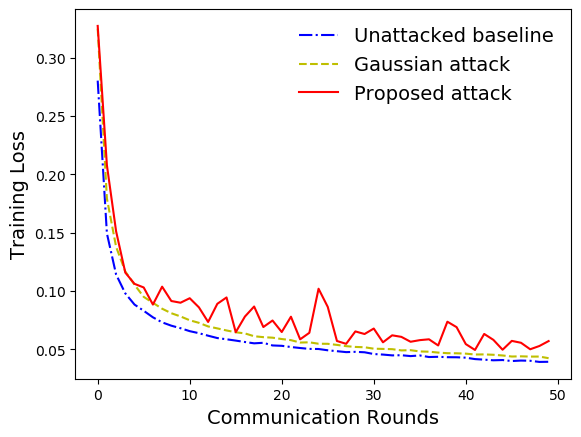}
    \label{fig:attack_krum_cnn}}
    \vspace{0.27cm}
    \subfloat[Trimmed Mean]{\includegraphics[width=7.5cm]{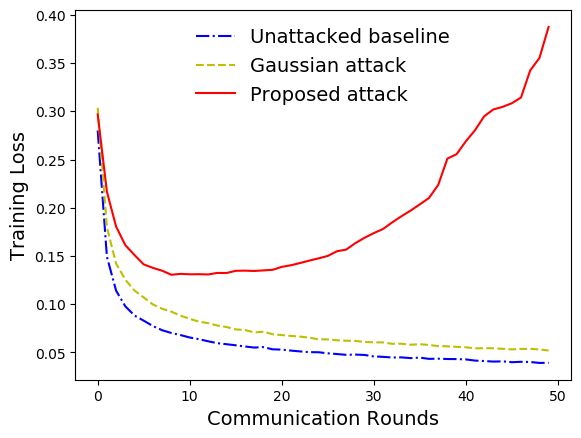}
    \label{fig:attack_trimmed_cnn}}
    \vspace{0.27cm}
    \caption{Model convergence for different attacks on CNN model.}
    \label{fig:attack_cnn}
\end{figure}

\begin{figure}
    \centering
    \subfloat[FedAvg]{\includegraphics[width=7.5cm]{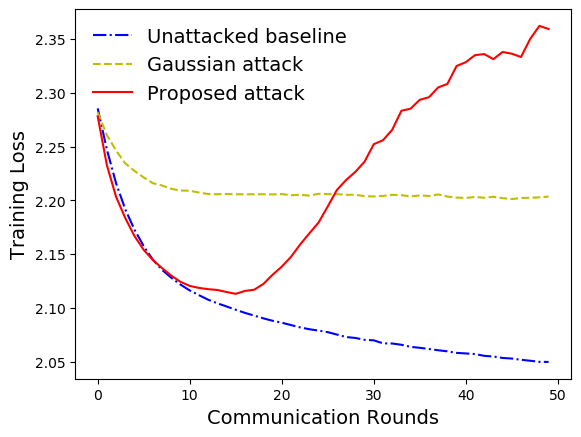}
    \label{fig:attack_avg_mlp}}
    \vspace{0.27cm}
    \subfloat[Krum]{\includegraphics[width=7.5cm]{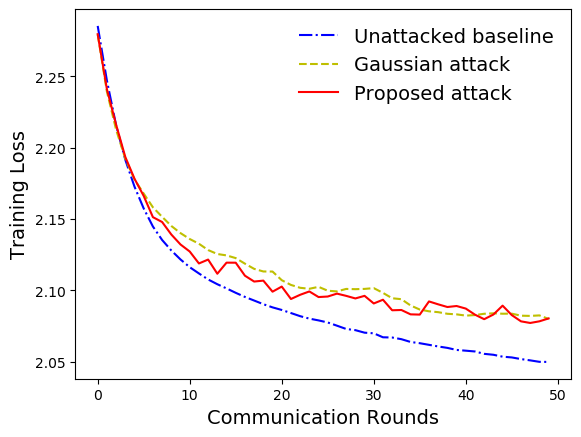}
    \label{fig:attack_krum_mlp}}
    \vspace{0.27cm}
    \subfloat[Trimmed Mean]{\includegraphics[width=7.5cm]{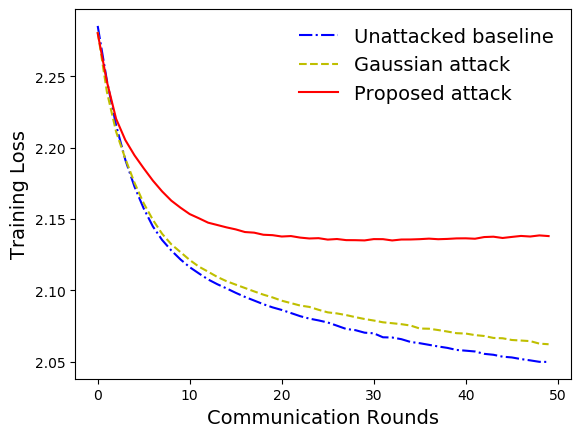}
    \label{fig:attack_trimmed_mlp}}
    \vspace{0.27cm}
    \caption{Model convergence for different attacks on MLP model.}
    \label{fig:attack_mlp}
\end{figure}

We also explore the model convergence under these attacks for three aggregation rules respectively. In Fig. \ref{fig:attack_cnn} and Fig. \ref{fig:attack_mlp}, it can be seen that our proposed attacks effectively slow down the model convergence when Krum is the aggregation rule in the server, and even lead the model to divergence in the presence of FedAvg and Trimmed Mean.

\subsection{Evaluation on Our Defenses}
We evaluate our proposed defense algorithm on MNIST and CIFAR-10 datasets using CNN and MLP models respectively. As shown in Table \ref{tab:def_cnn} and Table \ref{tab:def_mlp}, our defense is effective at optimizing the training loss for all scenarios. For example, when defending our proposed attack on the CNN model, the error rate remains at 3\%, which is the same as it when there is no attack. However, Krum and Trimmed Mean are not effective to defend our proposed attack. The error rates increase to 14\% and 85\% respectively. According to \eqref{er}, a small error rate reflects high testing accuracy, which means that the training loss achieves a local minimum after 50 rounds of training. This explains why our proposed defense method works, which eliminates the impacts of our proposed attack on both CNN and MLP models.

\begin{table}
    \centering
    \caption{Error Rates on CNN Model After Attacks for Defense Results}
    \label{tab:def_cnn}
    \begin{tabular}{|@{~}c@{~}|c|c|c|}
        \hline
        & No Attack & Gaussian Attack & \textbf{Proposed Attack} \\
        \hline
        Krum & 0.03 & 0.03 & 0.14 \\
        \hline
        Trimmed Mean & 0.03 & 0.05 & 0.85 \\
        \hline
        \textbf{Proposed Defense} & \textbf{0.03} & \textbf{0.03} & \textbf{0.03} \\
        \hline
    \end{tabular}
\end{table}
\begin{table}
    \centering
    \caption{Error Rates on MLP Model After Attacks for Defense Results}
    \label{tab:def_mlp}
    \begin{tabular}{|@{~}c@{~}|c|c|c|}
        \hline
        & No Attack & Gaussian Attack & \textbf{Proposed Attack} \\
        \hline
        Krum & 0.59 & 0.61 & 0.63 \\
        \hline
        Trimmed Mean & 0.59 & 0.59 & 0.65 \\
        \hline
        \textbf{Proposed Defense} & \textbf{0.59} & \textbf{0.59} & \textbf{0.59} \\
        \hline
    \end{tabular}
\end{table}

\section{Conclusion and Future Work} \label{conclusion}
This paper comprehensively analyzes diverse attacks and defenses on federated learning. The application of differential privacy in the context of federated learning is user-level differentially private. In this paper, we evaluate the vulnerabilities of differential privacy based federated learning and explore possible defense mechanism.

Our work raises the interests in the research direction of differential privacy based federated learning. We also explore the attacks and defense mechanisms with untargeted model poisoning attacks. In future research, targeted model poisoning attacks that are strongly related to data poisoning attacks will be investigated in real-world application scenarios such as pathologic diagnosis and intelligent agriculture.


\ifCLASSOPTIONcaptionsoff
  \newpage
\fi


\bibliographystyle{IEEEtran}
\bibliography{IEEEabrv,references}

\end{document}